\documentstyle[12pt,epsf]{article}

\begin{document}

\begin{center}

\underline{\sf To be published in: International Journal of Modern Physics C (2001) vol. 12 no. 2}

\vspace*{1cm}

{\large\bf Levy-stable distributions revisited:\\
tail index $>2$ does not exclude the Levy-stable regime}

\vspace*{.6cm}

{\large Rafa{\l} Weron}

\vspace*{.2cm}

{\small
Hugo Steinhaus Center for Stochastic Methods,\\
Wroc{\l}aw University of Technology, 50-370 Wroc{\l}aw, Poland\\
E-mail: rweron@im.pwr.wroc.pl
}

\end{center}

\vspace*{.5cm}

\noindent
{\bf Abstract:}
Power-law tail behavior and the summation scheme of Levy-stable distributions is the basis 
for their frequent use as models when fat tails above a Gaussian distribution are observed. 
However, recent studies suggest that financial asset returns exhibit tail exponents well 
above the Levy-stable regime ($0<\alpha\le 2$). In this paper we illustrate that widely 
used tail index estimates (log-log linear regression and Hill) can give exponents well
above the asymptotic limit for $\alpha$ close to 2, resulting in overestimation of the 
tail exponent in finite samples. The reported value of the tail exponent $\alpha$ 
around 3 may very well indicate a Levy-stable distribution with $\alpha\approx 1.8$.

\vspace*{.3cm}


\vspace*{.5cm}


\section{Introduction}

Levy-stable laws are a rich class of probability distributions that allow skewness 
and fat tails and have many intriguing mathematical properties \cite{STABLE}. 
They have been proposed as models for many types of physical and economic systems. 
There are several reasons for using Levy-stable laws to describe complex systems. 
First of all, in some cases there are solid theoretical reasons for expecting a non-Gaussian 
Levy-stable model, e.g. reflection off a rotating mirror yields a Cauchy distribution 
($\alpha=1$), hitting times for a Brownian motion yield a Levy distribution ($\alpha=0.5, 
\beta=1$), the gravitational field of stars yields the Holtsmark distribution ($\alpha=1.5$) 
\cite{feller71,zolotariev86,jw94}. 
The second reason is the Generalized Central Limit Theorem which states that the only 
possible non-trivial limit of normalized sums of independent identically distributed terms 
is Levy-stable \cite{CLT}. It is argued that some observed quantities are the sum of many 
small terms -- asset prices, noise in communication systems, etc. -- and hence a Levy-stable 
model should be used to describe such systems. 
The third argument for modeling with Levy-stable distributions is empirical: many large data 
sets exhibit fat tails (or heavy tails, as they are called in the mathematical literature) 
and skewness, for a review see \cite{jw94,ekm97,rm00}. Such data sets are poorly described 
by a Gaussian model and usually can be quite well described by a Levy-stable distribution.

Recently, in a series of economic and econophysics articles the Levy-stability of returns 
has been rejected based on the log-log linear regression of the cumulative distribution 
function or the Hill estimator 
\cite{dumouchel83,hv91,lp94,longin96,dv97,gmas98,gpams99,pgams99,pgras00,la01}. 
In this paper we show that the cited estimation methods can give exponents well
above the asymptotic limit for Levy-stable distributions with $\alpha$ close to 2, 
which results in overestimation of the tail exponent in finite samples. As a consequence, 
the reported value of the tail exponent $\alpha$ around 3 may suggest a Levy-stable 
distribution with $\alpha\approx 1.8$.

\section{Levy-stable distributions}

Levy-stable laws were introduced by Paul Levy during his investigations of the behavior 
of sums of independent random variables in the early 1920's \cite{levy25}.
The lack of closed form formulas for probability density functions for all but three Levy-stable
distributions (Gaussian, Cauchy and Levy), has been a major drawback to the use of Levy-stable 
distributions by practitioners. However, now there are reliable computer programs to compute 
Levy-stable densities, distribution functions and quantiles \cite{nolan97}. With these programs, 
it is possible to use Levy-stable models in a variety of practical problems. 

The Levy-stable distribution requires four parameters to describe: an index of stability 
(tail index, tail exponent or characteristic exponent) $\alpha\in (0,2]$, a skewness parameter 
$\beta\in [-1,1]$, a scale parameter $\sigma>0$ and a location parameter $\mu\in R$. 
The tail exponent $\alpha$ determines the rate at which the tails of the distribution 
taper off, see Fig. 1. When $\alpha=2$, a Gaussian distribution results. When $\alpha<2$, 
the variance is infinite. When $\alpha>1$, the mean of the distribution exists and is equal 
to $\mu$. In general, the $p$--th moment of a Levy-stable random variable is finite if and 
only if $p<\alpha$. When the skewness parameter $\beta$ is positive, the distribution is 
skewed to the right. When it is negative, it is skewed to the left. When $\beta=0$, the 
distribution is symmetric about $\mu$. As $\alpha$ approaches 2, $\beta$ loses its effect 
and the distribution approaches the Gaussian distribution regardless of $\beta$.
The last two parameters, $\sigma$ and $\mu$, are the usual scale and location parameters, 
i.e. $\sigma$ determines the width and $\mu$ the shift of the mode (the peak) of the distribution.

\subsection{Characteristic function representation}

Due to the lack of closed form formulas for densities, the Levy-stable distribution can be 
most conveniently described by its characteristic function $\phi(t)$ -- the inverse Fourier 
transform of the probability density function. 
However, there are multiple parameterizations for Levy-stable laws and much confusion has been 
caused by these different representations \cite{hall81}. The variety of formulas is 
caused by a combination of historical evolution and the numerous problems that have been 
analyzed using specialized forms of the Levy-stable distributions. The most popular 
parameterization of the characteristic function of $X \sim S_\alpha(\sigma,\beta,\mu)$, 
i.e. a Levy-stable random variable with parameters $\alpha$, $\sigma$, $\beta$ and $\mu$, 
is given by \cite{st94,weron96}:
\begin{equation}\label{def-stab-a}
\log\phi(t) = \cases{
 -\sigma^{\alpha}|t|^{\alpha}\{1-i\beta {\rm sign}(t)\tan\frac{\pi\alpha}{2}\}+i\mu t, 
   & $\alpha \ne 1,$ \cr\cr
 -\sigma|t|\{1+i\beta {\rm sign}(t)\frac{2}{\pi}\log|t|\}+i\mu t, 
   & $\alpha=1.$}
\end{equation}
For numerical purposes, it is often useful \cite{fn99} to use a different parameterization:
\begin{equation}\label{def-stab-m}
\log\phi_0(t) = \cases{
 -\sigma^{\alpha}|t|^{\alpha}\{1+i\beta {\rm sign}(t)\tan\frac{\pi\alpha}{2}[(\sigma |t|)^{1-\alpha}-1]\}+i\mu_0 t, 
   & $\alpha \ne 1,$ \cr\cr
 -\sigma|t|\{1+i\beta {\rm sign}(t)\frac{2}{\pi}\log(\sigma |t|)\}+i\mu_0 t, 
   & $\alpha=1.$}
\end{equation}
The $S^0_{\alpha}(\sigma,\beta,\mu_0)$ parameterization is a variant of Zolotariev's 
\cite{zolotariev86} (M)-parameterization, with the characteristic function and hence 
the density and the distribution function jointly continuous in all four parameters. 
In particular, percentiles and convergence to the power-law tail vary in a continuous 
way as $\alpha$ and $\beta$ vary. The location parameters of the two representations 
are related by $\mu = \mu_0 - \beta\sigma\tan\frac{\pi\alpha}{2}$ for $\alpha\ne 1$ 
and $\mu = \mu_0 - \beta\sigma\frac{2}{\pi}\log\sigma$ for $\alpha=1$.

For simplicity, in Section 3 we will analyze only non-skewed ($\beta=0$) Levy-stable laws. 
This is not a very restrictive assumption, since most financial asset returns exhibit 
only slight skewness. For $\beta=0$ both representations are equivalent, however, in 
the general case the $S^0$ representation is preferred.

\begin{figure}[p]
\centerline{\epsfxsize=10cm \epsfbox{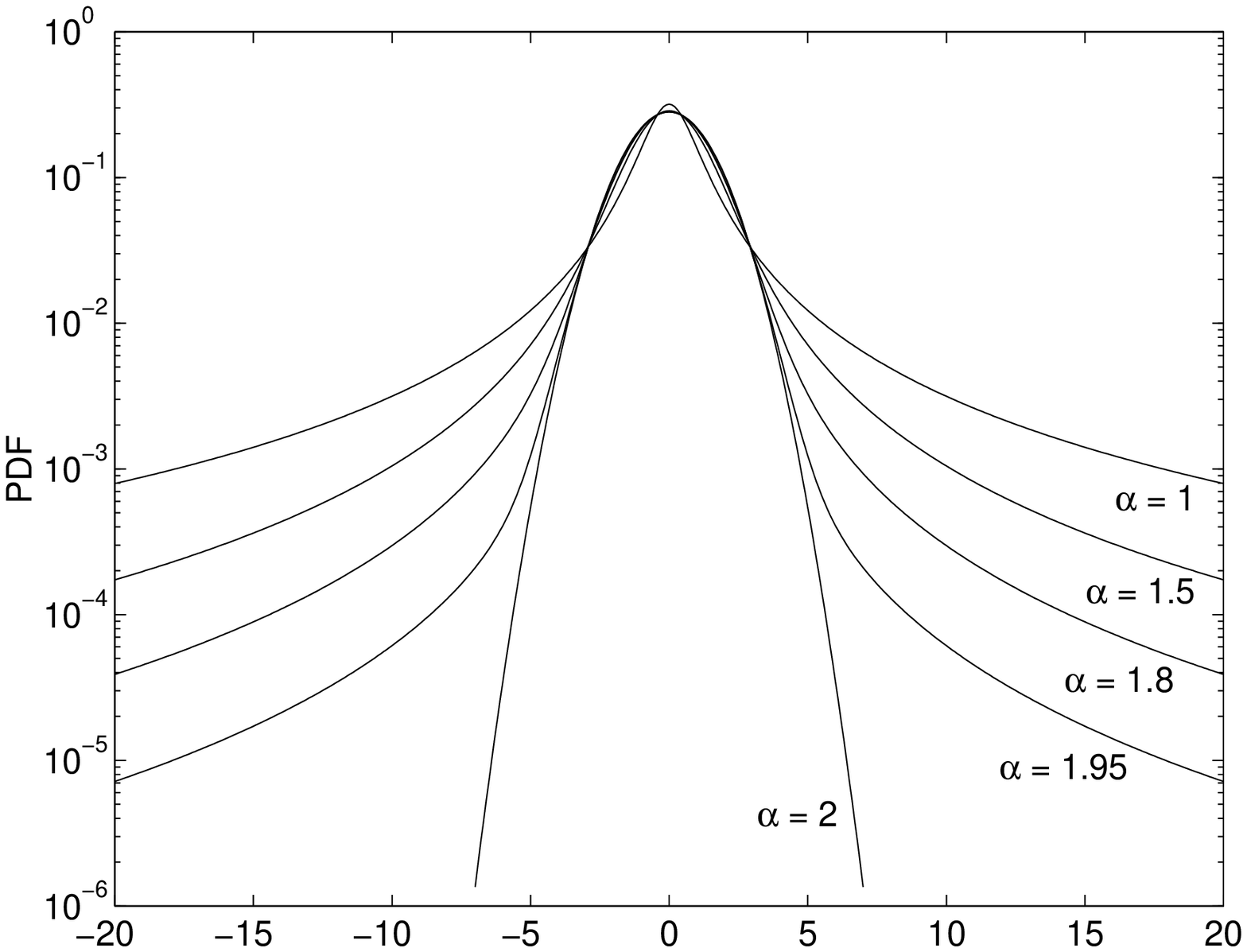}}
\caption{A semilog plot of symmetric ($\beta=\mu=0$) Levy-stable probability density functions 
for $\alpha = 2, 1.95, 1.8, 1.5$ and $1$. Observe that the Gaussian ($\alpha=2$) density 
forms a parabola and is the only Levy-stable density with exponential tails.}

\centerline{\epsfxsize=10cm \epsfbox{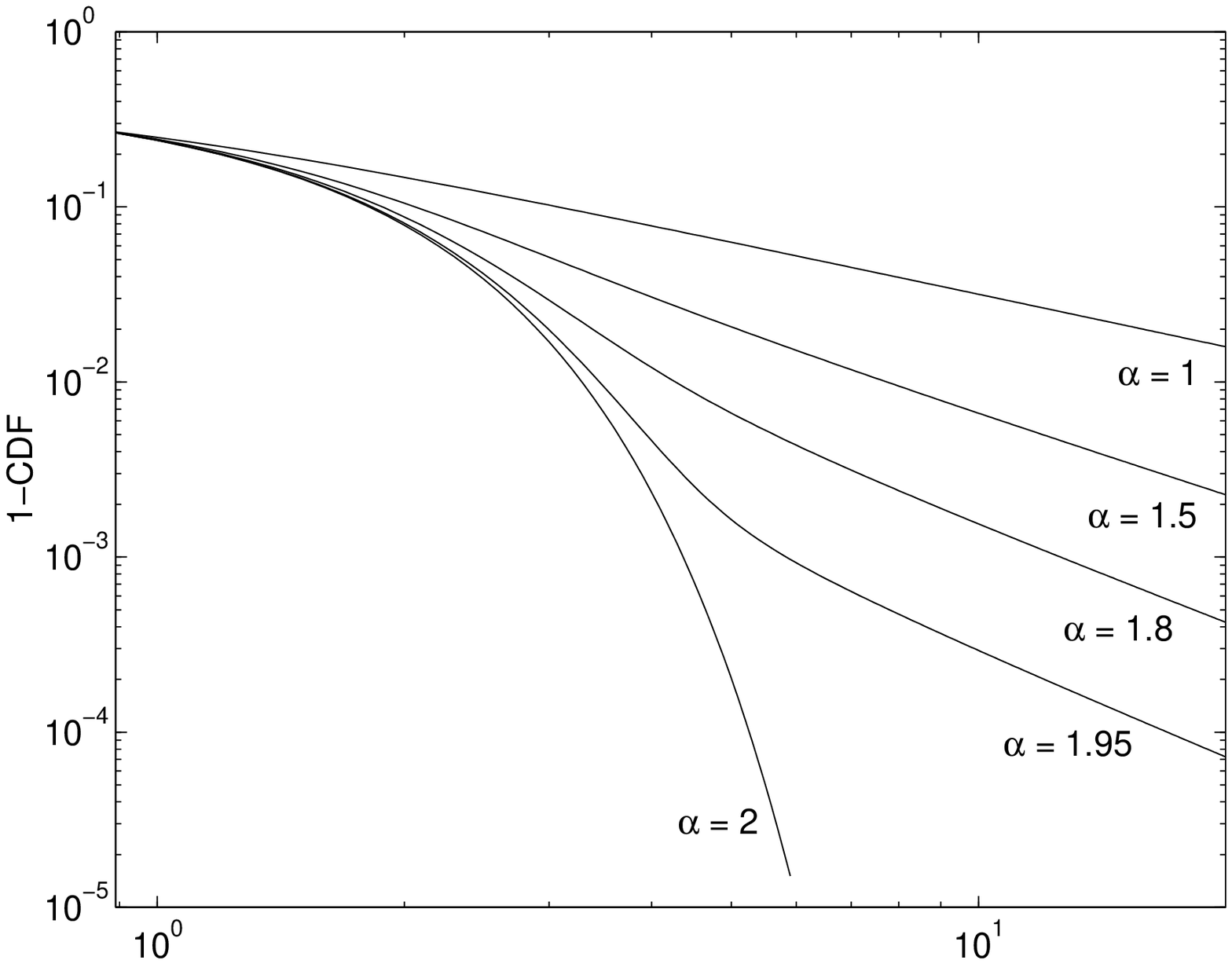}}
\caption{A double logarithmic plot of the right tails of symmetric Levy-stable cumulative 
distribution functions for $\alpha = 2, 1.95, 1.8, 1.5$ and $1$. For $\alpha<2$, 
the power tails are clearly visible. Moreover, the smaller the tail index the stronger is 
the power decay behavior. Recall that the Gaussian tails decay much faster, i.e. exponentially.}
\end{figure}

\subsection{Simulation of Levy-stable variables}

The complexity of the problem of simulating sequences of Levy-stable random variables results 
from the fact that there are no analytic expressions for the inverse $F^{-1}$ of the 
cumulative distribution function. 
The first breakthrough was made by Kanter \cite{kanter75}, who gave a direct method for 
simulating $S_\alpha(1,1,0)$ random variables, for $\alpha<1$. It turned out that this 
method could be easily adapted to the general case. Chambers, Mallows and Stuck \cite{cms76} 
were the first to give the formulas.

The algorithm for constructing a random variable $X\sim S_\alpha(1,\beta,0)$, in representation 
(\ref{def-stab-a}), is the following \cite{weron96}:
\begin{itemize}
\item 
generate a random variable $V$ uniformly distributed on $(-\frac{\pi}{2},\frac{\pi}{2})$ 
and an independent exponential random variable $W$ with mean 1;
\item 
for $\alpha\ne 1$ compute:
\begin{equation}\label{gen-skewstab}
X = S_{\alpha,\beta} \times \frac{\sin(\alpha(V+B_{\alpha,\beta}))}{(\cos(V))^{1/\alpha}} 
    \times \left(\frac{\cos(V - \alpha(V+B_{\alpha,\beta}))}{W}\right)^{(1-\alpha)/\alpha},
\end{equation}
where
\begin{eqnarray*}
B_{\alpha,\beta} & = & \frac{\arctan(\beta\tan \frac{\pi\alpha}{2})}{\alpha}, \\
S_{\alpha,\beta} & = & \left[1+\beta^2 \tan^2\frac{\pi\alpha}{2}\right]^{1/(2\alpha)};
\end{eqnarray*}
\item 
for $\alpha=1$ compute:
\begin{equation}\label{gen-skewstab-1}
X = \frac{2}{\pi}\left[ \left(\frac{\pi}{2}+\beta V \right)\tan V-
 \beta\log\left(\frac{\frac{\pi}{2} W\cos V}{\frac{\pi}{2}+\beta V}\right)\right].
\end{equation}
\end{itemize}
Given the formulas for simulation of standard Levy-stable random variables, we can easily
simulate a Levy-stable random variable for all admissible values of the parameters $\alpha$, 
$\sigma$, $\beta$ and $\mu$ using the following property: if $X\sim S_\alpha(1,\beta,0)$ then
\begin{eqnarray*}
Y=\cases{
 \sigma X+\mu, & $\alpha \ne 1$, \cr\cr
 \sigma X+\frac{2}{\pi}\beta\sigma\log\sigma + \mu, & $\alpha=1$,}
\end{eqnarray*} 
is $S_\alpha(\sigma,\beta,\mu)$. The presented method is regarded as the fastest and the best 
one known. It is widely used in many software packages, including S-plus and STABLE \cite{nolan97}.

\subsection{Tail behavior}

Levy \cite{levy25} has shown that when $\alpha<2$ the tails of Levy-stable distributions are 
asymptotically equivalent to a Pareto law. Namely, if $X \sim S_{\alpha<2}(1,\beta,0)$ then 
as $x \rightarrow \infty$:
\begin{eqnarray}\label{stab-tail}
P(X>x) = 1-F(x) & \rightarrow & C_\alpha (1+\beta) x^{-\alpha}, \nonumber \\
& & \\
P(X<-x) = F(-x) & \rightarrow & C_\alpha (1-\beta) x^{-\alpha}, \nonumber
\end{eqnarray}
where 
$$
C_\alpha 
= \left(2 \int_0^\infty x^{-\alpha} \sin x dx \right)^{-1} 
= \frac{1}{\pi} \Gamma(\alpha) \sin \frac{\pi\alpha}{2}.
$$
The convergence to a power-law tail varies for different $\alpha$'s and, as can be seen 
in Fig. 2, is slower for larger values of the tail index. Moreover, the tails of Levy-stable 
distribution functions exhibit a crossover from a power decay with exponent $\alpha>2$ 
to the true tail with exponent $\alpha$. This phenomenon is more visible for large 
$\alpha$'s and will be investigated further in the next Section.

\section{Estimation of the tail index}

The problem of estimating the tail index (as well as other parameters) is in 
general severely hampered by the lack of known closed--form density functions for all 
but a few members of the Levy-stable family. Fortunately, there are numerical methods 
that have been found useful in practice.

Currently there exist three estimation procedures for estimating Levy-stable law parameters
worth recalling. The first one is based on a numerical approximation of the Levy-stable 
likelihood function. The ML method, as it is called, was originally developed by DuMouchel 
\cite{dumouchel73} and recently optimized by Nolan \cite{nolan99}. It is the slowest of the
three but possesses well known asymptotic properties.

The second method uses a regression on the sample characteristic function. It is both fast 
and accurate (as long as we are dealing with a sample generated by a Levy-stable law). The
regression procedure was developed in the early 1980's by Koutrouvelis 
\cite{koutrouvelis80,koutrouvelis81} and recently improved by Kogon and Williams \cite{kw98}.

The last, but not least, is the quantile method of McCulloch \cite{mcculloch86}. It is the 
fastest of the three, because it is based on tabulated quantiles of Levy-stable laws. 
Yet it lacks the universality of the other two, since it is restricted to $\alpha\ge 0.6$.

All presented methods work pretty well assuming that the sample under consideration
is indeed Levy-stable. However, if the data comes from a different distribution, these 
procedures may mislead more then the Hill and direct tail estimation methods. And since 
there are no formal tests for assessing the Levy-stability of a data set we suggest to first 
apply the "visual inspection" or non-parametric tests to see whether the empirical 
densities resemble those of Levy-stable laws.

\subsection{Log-log linear regression}

The simplest and most straightforward method of estimating the tail index is to plot 
the right tail of the (empirical) cumulative distribution function (i.e. $1-F(x)$) 
on a double logarithmic paper, as in Figs. 2-7 (see also Refs. 
\cite{gmas98,gpams99,pgams99,pgras00}). 
The slope of the linear regression for large values of $x$ yields the estimate of 
the tail index $\alpha$, through the relation $\alpha = -\mbox{slope}$.

This method is very sensitive to the sample size and the choice of the number of observations 
used in the regression. Moreover, the slope around $-3$ may indicate a non-Levy-stable power-law 
decay in the tails or the contrary -- a Levy-stable distribution with $\alpha\approx 1.8$.
To illustrate this, we simulated (using Eq. (\ref{gen-skewstab})) samples of size 
$N = 10^4$ and $10^6$ of standard symmetric ($\beta=\mu=0$, $\sigma=1$) Levy-stable distributed 
variables with $\alpha=1.95$ and $1.8$. Next, we plotted the right tails of the empirical 
distribution functions on a double logarithmic paper. For $\alpha$ close to 2 the true tail 
behavior (\ref{stab-tail}) is observed only for very large (also for very small, i.e. the 
negative tail) observations, see Figs. 3 and 5, after a crossover from a temporary power-like 
decay. Moreover, the obtained estimates still have a slight positive bias: $0.03$ for 
$\alpha=1.8$ and $0.12$ for $\alpha=1.95$, which suggests that perhaps even larger samples
than $10^6$ observations should be used. 

To test the method for extremely large data sets we also simulated samples of size $N = 10^8$ 
of standard symmetric Levy-stable distributed variables with $\alpha=1.95$ and $1.8$. As can
be seen in Fig. 7 the improvement over one million records samples is not substantial. 
The tail index estimate (2.02) is closer to the true value of $\alpha=1.95$, but still 
outside the Levy-stable regime. Similarly, for $\alpha=1.8$ the method overestimated the 
tail exponent and returned $\alpha=1.82$.

If a typical size data set is used, i.e. $10^4$ observations or less as in Refs. \cite{lp94,la01} 
and some data sets in Refs. \cite{gpams99,pgams99,pgras00}, the plot may be quite misleading. 
An empirical cumulative distribution function of a Levy-stable sample of size $10^4$ and 
$\alpha>1.5$ does not exhibit the true tail behavior ($x^{-\alpha}$ decay), but a temporary 
power-like decay with the slope (more precisely: absolute value of the slope) significantly 
greater then 2: $|\mbox{slope}|$ $=4.99$ for $\alpha=1.95$ and 2.89 for $\alpha=1.8$; 
see Figs. 4 and 6 and compare with Figs. 3 and 5, respectively. Slight differences in 
the slope of the initial temporary power-like decay (4.91 in Fig. 3 compared to 4.99 in Fig. 4
and 2.95 in Fig. 5 compared to 2.89 in Fig. 6) are caused by inaccurate estimation for 
large observations in the smaller samples.

Figures 3-7 clearly illustrate that the true tail behavior of Levy-stable laws is visible only
for extremely large data sets. In practice, this means that in order to estimate $\alpha$ we 
must use high-frequency asset returns and restrict ourselves to the most "outlying" observations. 
Otherwise, inference of the tail index may be strongly misleading and rejection of the 
Levy-stable regime unfounded. 
In Figures 3, 5 and 7 we used only the upper 0.5\% or less of the records to estimate the true 
tail exponent. In general, the choice of observations used in the regression is subjective 
and can yield large estimation errors. 

\begin{figure}[p]
\centerline{\epsfxsize=10cm \epsfbox{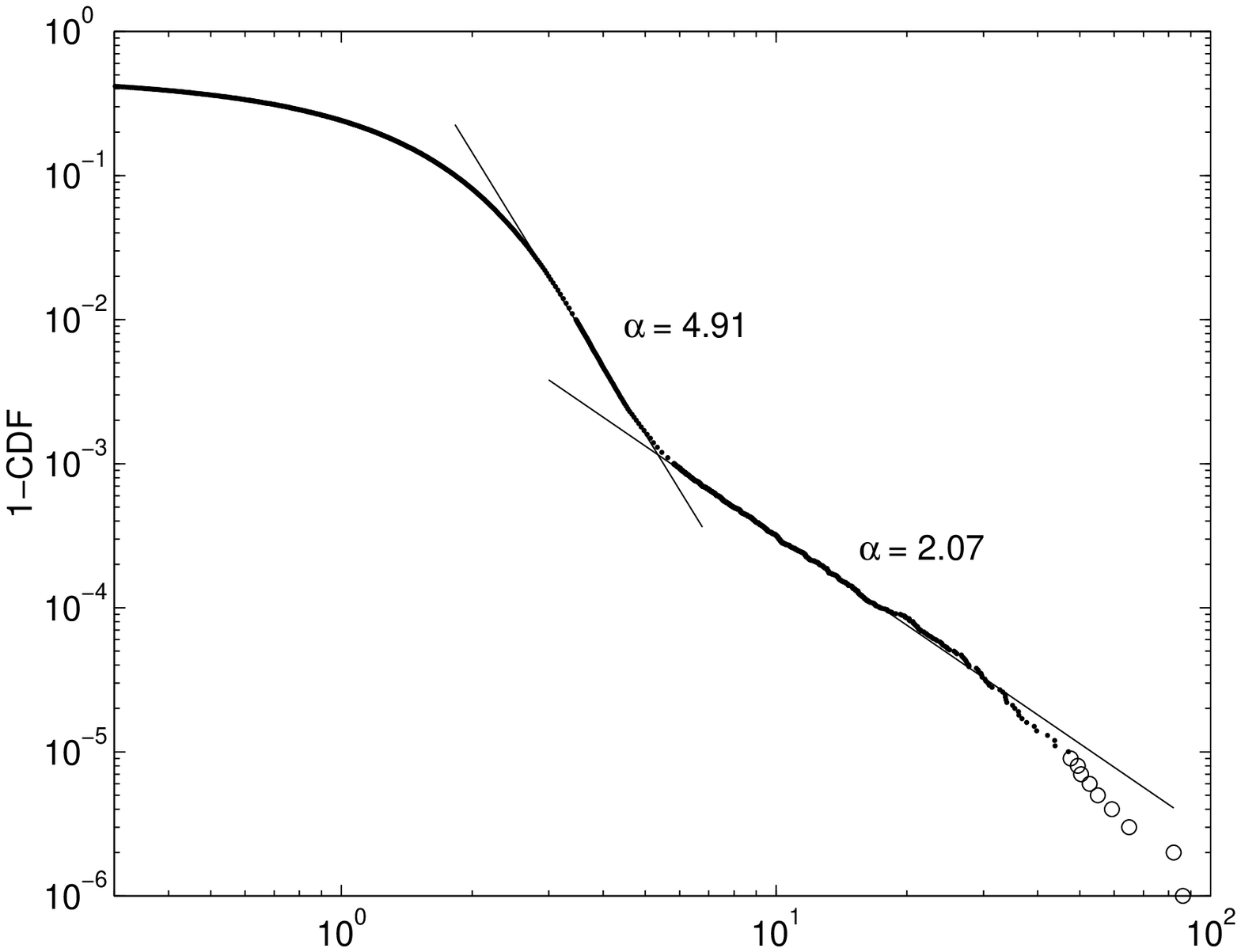}}
\caption{A double logarithmic plot of the right tail of an empirical symmetric $1.95$-stable 
distribution function for sample size $N = 10^6$. Circles represent outliers which were not 
used in the estimation process. Even the far tail estimate $\alpha=2.07$ is above 
the Levy-stable regime.}

\centerline{\epsfxsize=10cm \epsfbox{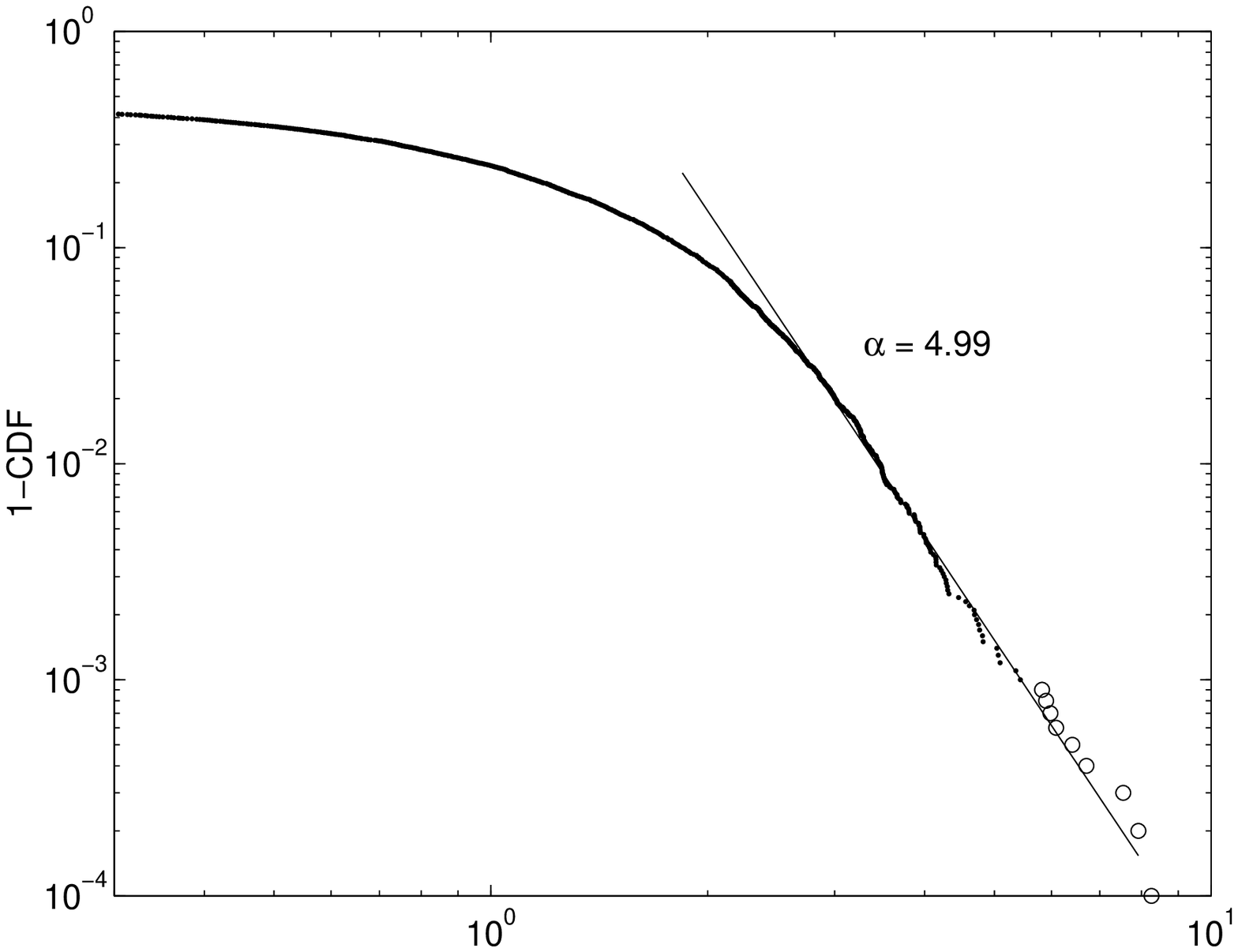}}
\caption{A double logarithmic plot of the right tail of an empirical symmetric $1.95$-stable 
distribution function for sample size $N = 10^4$. Circles represent outliers which were not 
used in the estimation process. This example shows that inference of the tail exponent from 
samples of typical size is strongly biased.}
\end{figure}

\begin{figure}[p]
\centerline{\epsfxsize=10cm \epsfbox{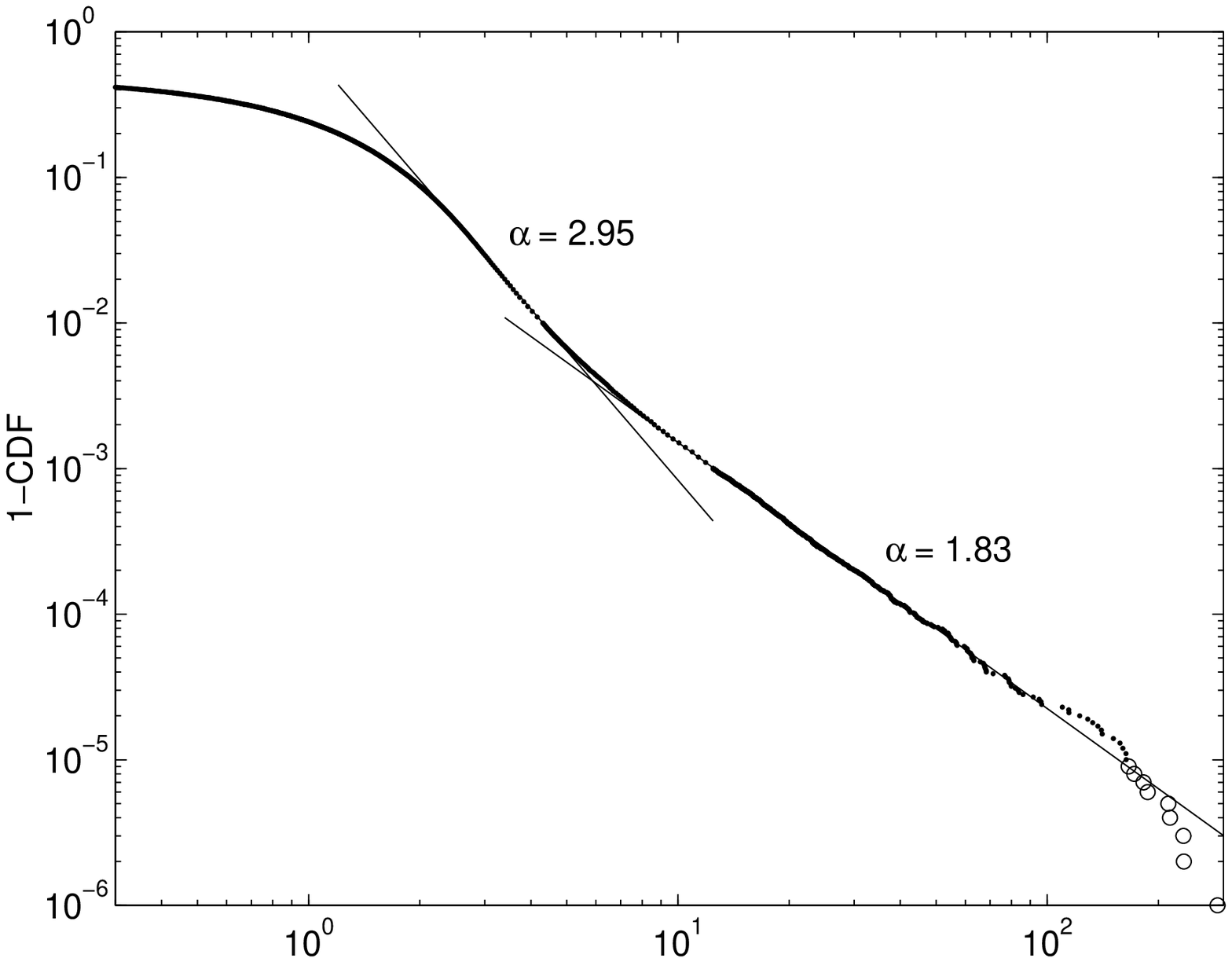}}
\caption{A double logarithmic plot of the right tail of an empirical symmetric $1.8$-stable 
distribution function for sample size $N = 10^6$. Circles represent outliers which were not 
used in the estimation process. The far tail estimate $\alpha=1.83$ is slightly 
above the true value of $\alpha$.}

\centerline{\epsfxsize=10cm \epsfbox{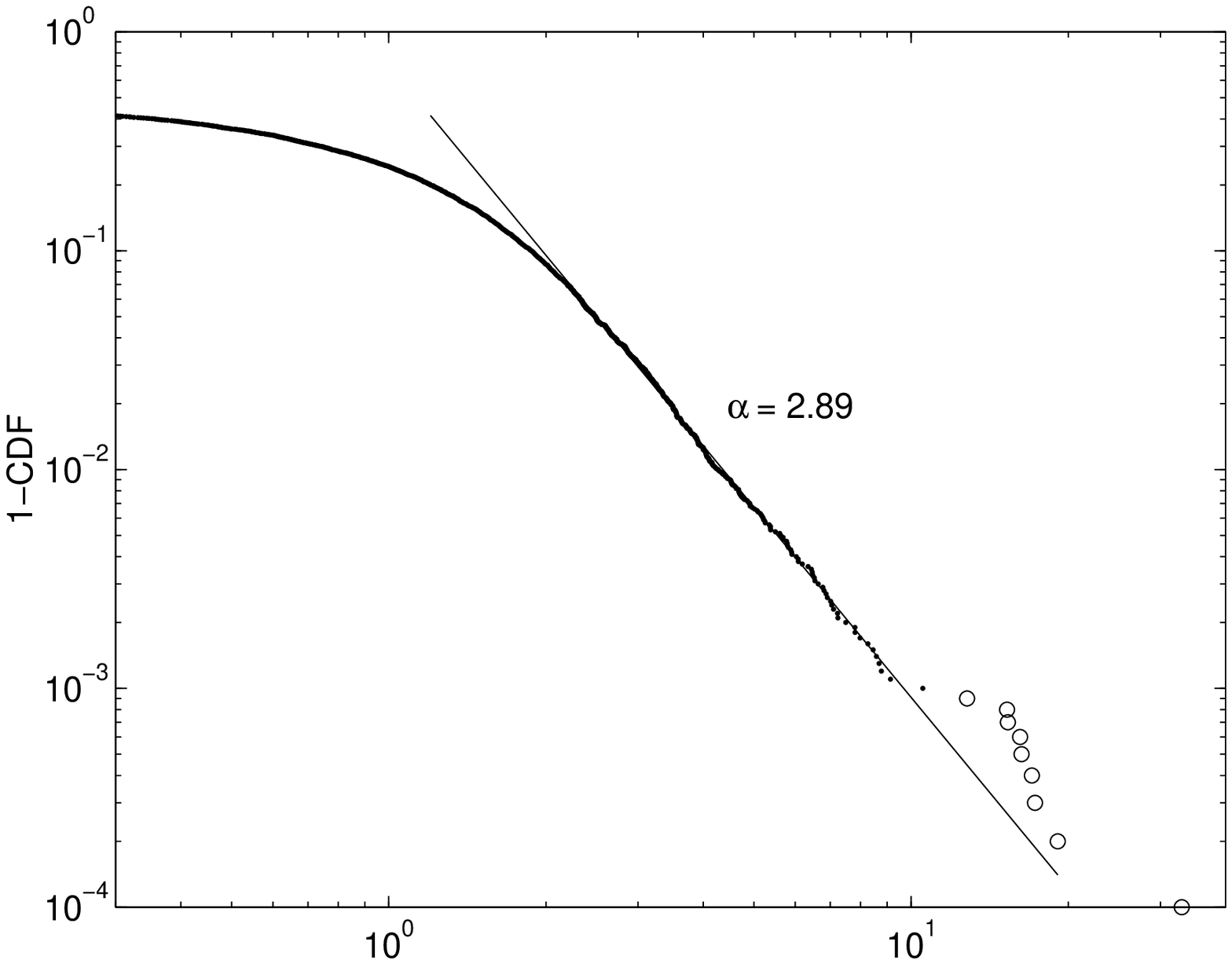}}
\caption{A double logarithmic plot of the right tail of an empirical symmetric $1.8$-stable 
distribution function for sample size $N = 10^4$. Circles represent outliers which were not 
used in the estimation process. This example shows that inference of the tail exponent from 
samples of typical size is strongly biased and the reported value of the tail exponent  
around 3 may very well indicate a Levy-stable distribution with $\alpha\approx 1.8$.}
\end{figure}

\begin{figure}[p]
\centerline{\epsfxsize=10cm \epsfbox{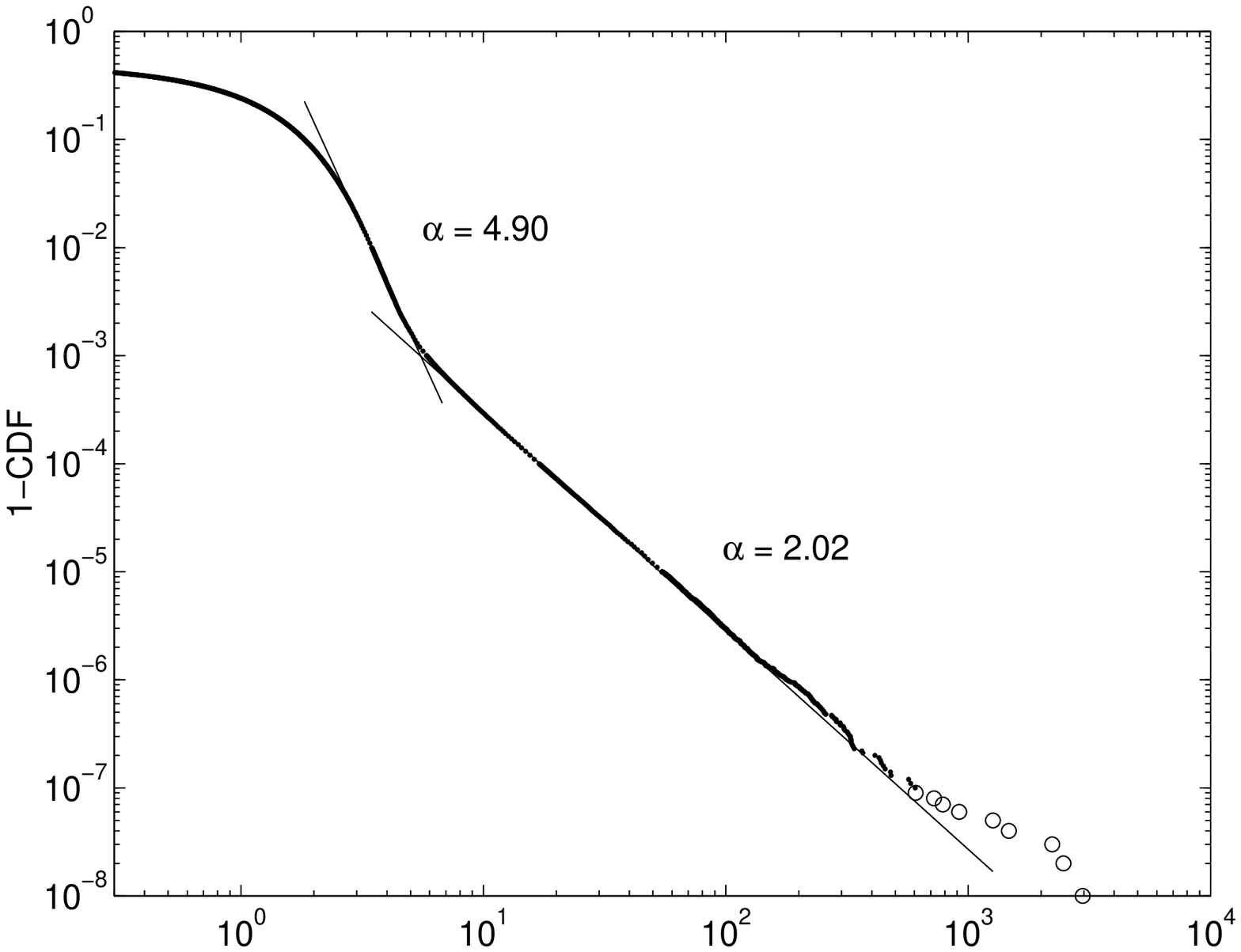}}
\caption{A double logarithmic plot of the right tail of an empirical symmetric $1.95$-stable 
distribution function for sample size $N = 10^8$. Circles represent outliers which were not 
used in the estimation process. Even the far tail estimate $\alpha=2.02$ is above 
the Levy-stable regime.}

\centerline{\epsfxsize=10cm \epsfbox{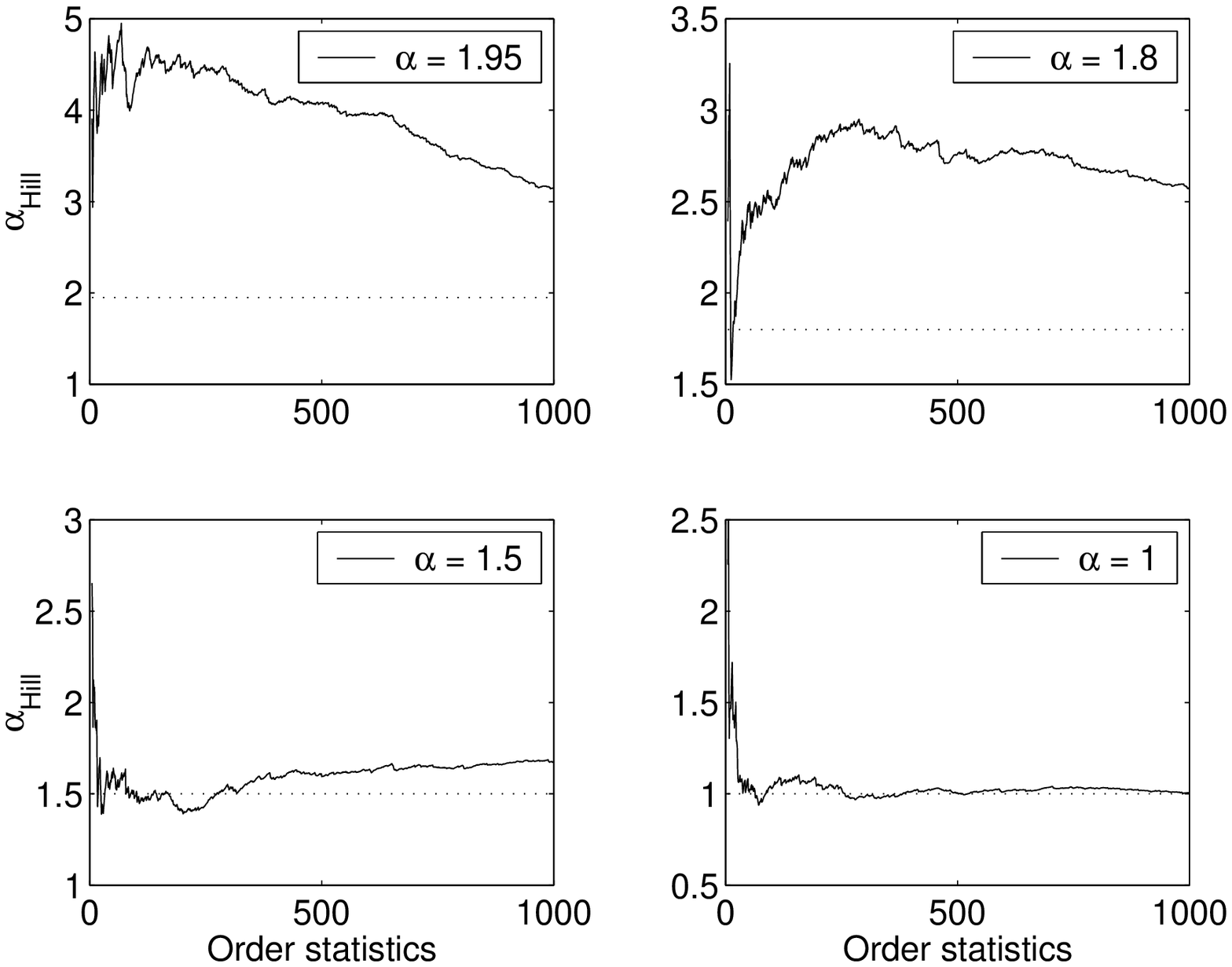}}
\caption{Plots of the Hill statistics $\alpha_{Hill}$ vs. the maximum order statistic $k$
for $1.95, 1.8, 1.5$ and $1$-stable samples of size $N = 10^4$. Dashed lines represent 
the true value of $\alpha$. For $\alpha$ close to 2, the Hill tail estimator has 
a large positive bias resulting in overestimation of the tail exponent.}
\end{figure}

\begin{figure}[p]
\centerline{\epsfxsize=10cm \epsfbox{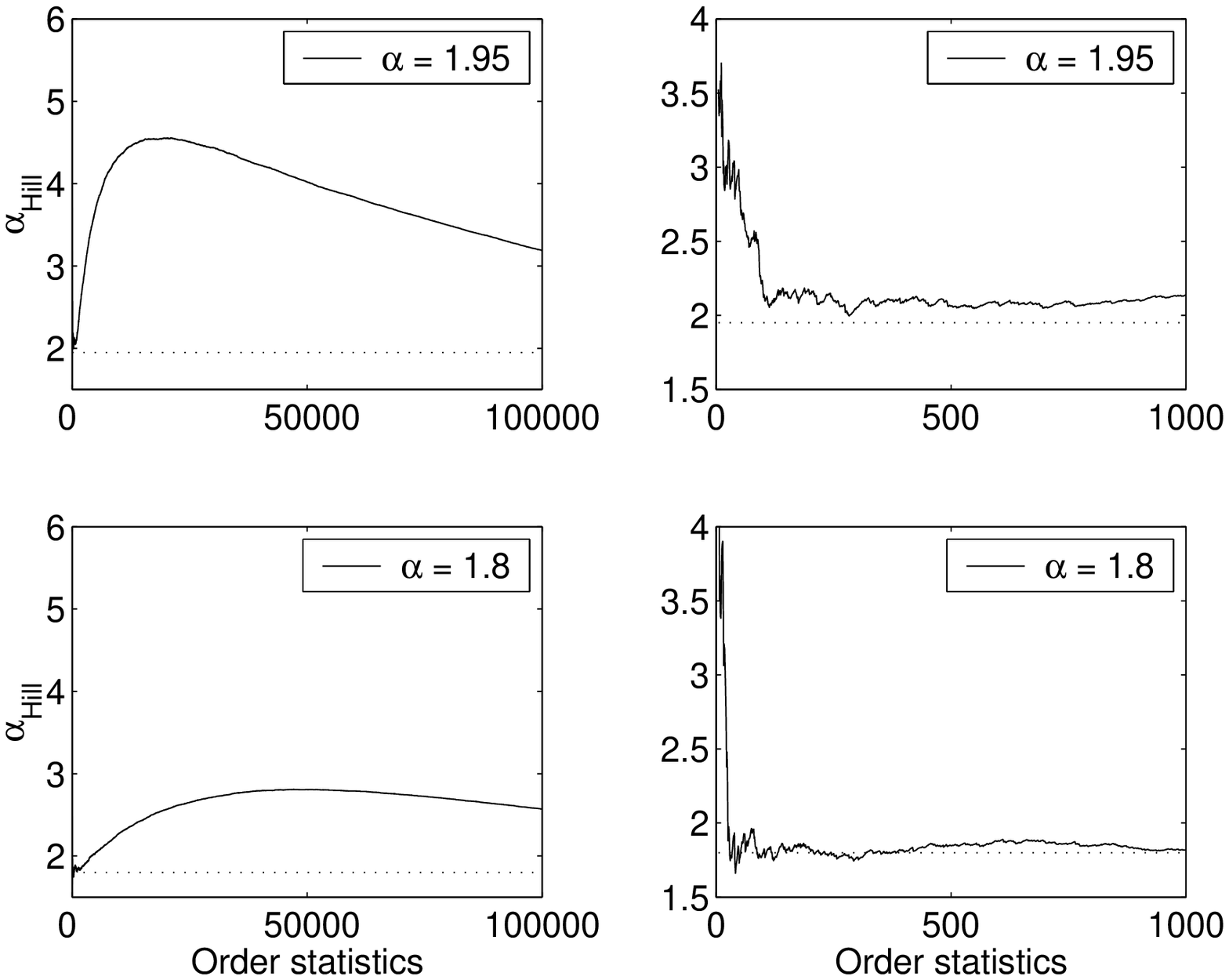}}
\caption{Plots of the Hill statistics $\alpha_{Hill}$ vs. the maximum order statistic $k$
for $1.95$ and $1.8$-stable samples of size $N = 10^6$. Dashed lines represent 
the true value of $\alpha$. For better visibility, right plots are a magnification of the 
left plots for small $k$. For $\alpha=1.8$ a good estimate is obtained only for $k=50,...,400$
(i.e. for $k<0.04\%$ of sample size), whereas for $\alpha=1.95$ the estimate is always above 
the Levy-stable regime.}

\centerline{\epsfxsize=10cm \epsfbox{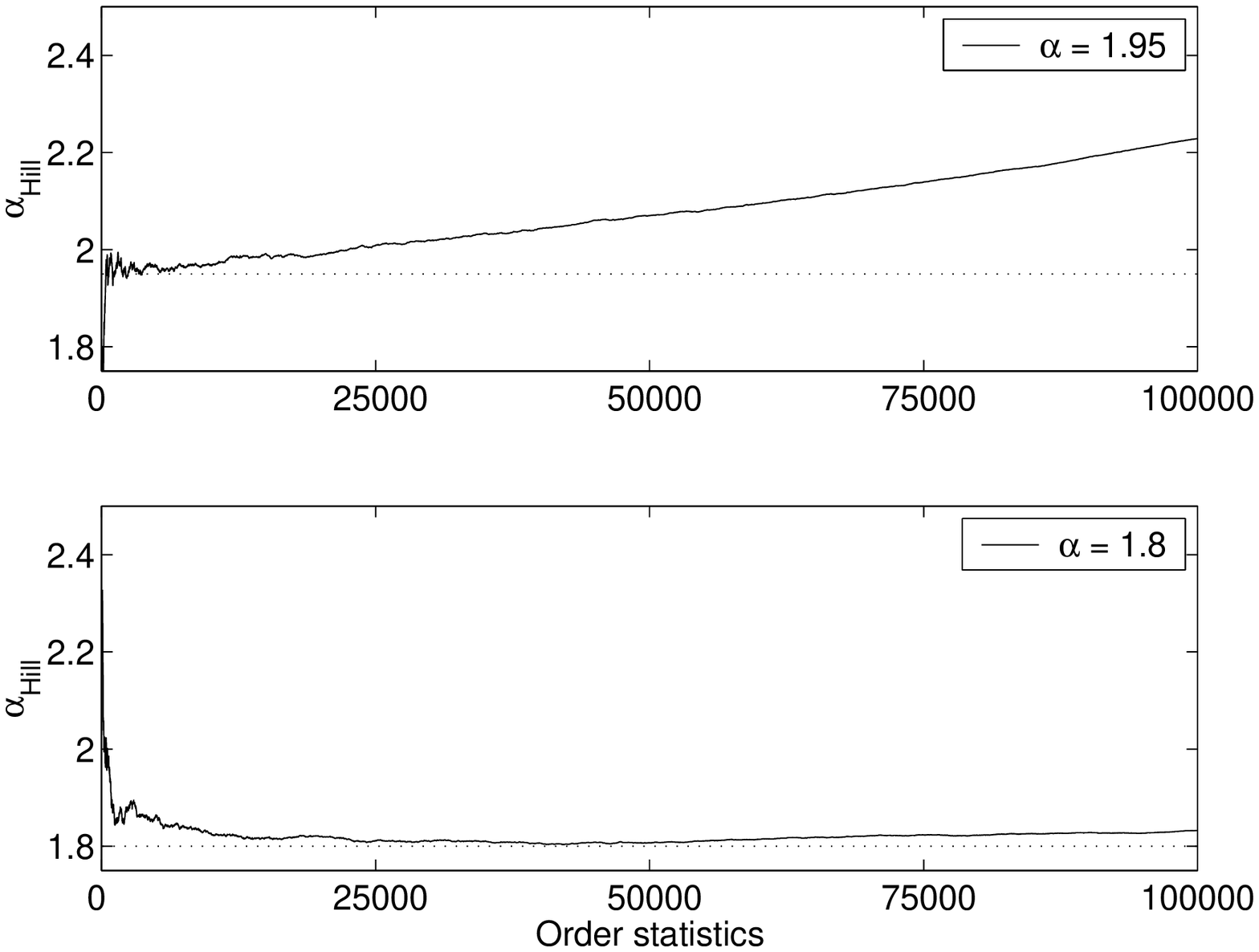}}
\caption{Plots of the Hill statistics $\alpha_{Hill}$ vs. the maximum order statistic $k$
for $1.95$ and $1.8$-stable samples of size $N = 10^8$. Dashed lines represent the true value 
of $\alpha$. For $\alpha=1.8$ a good estimate is obtained only for $k=25000,...,60000$
(i.e. for $k<0.06\%$ of sample size), whereas for $\alpha=1.95$ the estimate is in the 
Levy-stable regime only for $k<20000$ (i.e. for $k<0.02\%$ of sample size).}
\end{figure}

\subsection{Hill estimator}

Hill \cite{hill75} proposed a method for estimating the tail index that does not assume
a parametric form for the entire distribution function, but focuses only on the 
tail behavior. The Hill estimator is used to estimate the (Pareto) tail index $\alpha$,
when the upper tail \cite{HILL} of the distribution is of the form: $1-F(x) = C x^{-\alpha}$.
If $X_{(1)}, X_{(2)}, ..., X_{(N)}$ is the order statistics, i.e. original sample ordered so 
that $X_{(1)} \ge X_{(2)} \ge ... \ge X_{(N)}$, drawn from a population with law $F$ then 
the Hill estimate of $\alpha$ based on the $k$ largest order statistics is:
\begin{equation}\label{eqn-hill}
\alpha_{Hill}(k) = \left( \frac1k \sum_{n=1}^k \log \frac{X_{(n)}}{X_{(k+1)}} \right)^{-1}.
\end{equation}
Unfortunately, it is difficult to choose the right value of $k$. In practice,
$\alpha_{Hill}(k)$ is plotted against $k$ and one looks for a region where the plot levels off
to identify the correct order statistic \cite{ekm97,resnick97}. Algorithms for choosing
"optimal" $k$ have been proposed in the literature (see e.g. Refs. \cite{bvt96,dk98}), 
but usually give numbers in the vicinity of the plateau.

To illustrate the performance of the Hill estimator for Levy-stable laws, we simulated 
(using Eq. (\ref{gen-skewstab})) samples of size $N = 10^4, 10^6$ and $10^8$ of standard 
symmetric ($\beta=\mu=0$, $\sigma=1$) Levy-stable distributed variables with $\alpha=1.95, 
1.8, 1.5$ and 1. Next, we plotted the Hill statistic $\alpha_{Hill}(k)$ vs. $k$ and 
compared estimated and true values of $\alpha$.

Figure 8 presents the results of the analysis for Levy-stable samples of size $10^4$. As has 
been reported in the literature \cite{resnick97,mcculloch97,rt97}, for $\alpha\le 1.5$ 
the estimation is within reasonable limits but as $\alpha$ approaches 2 the Hill estimate 
is well above the Levy-stable regime. In the "extreme" case of $\alpha=1.95$ there is no 
single value of $k$ that gives the right value of the tail exponent! For the $1.8$-stable 
sample Hill estimates are close to the true value of $\alpha$ only for $k<50$, whereas the 
plateau and "optimal" order statistic \cite{dk98} are in the range $k\in (200,400)$ yielding
$\alpha_{Hill}\approx 2.85$. Actually the results are very close to those of the log-log 
linear regression, see Figs. 4 and 6 and Table 1, since in both methods estimates are obtained
from the largest order statistics.

\begin{table}[htbp]
\caption{Pareto (power-law) tail index $\alpha$ estimates for Levy-stable samples of size $10^4$.
}
\begin{center}
\begin{tabular}{ccc}
\hline
 & \multicolumn{2}{c}{Estimation method} \\
Simulated $\alpha$ & Log-log regression & Hill \\
\hline
1.95 & 4.19 & 4.4 -- 4.6 \\
1.8  & 2.89 & 2.8 -- 2.9  \\
1.5  & ---  & 1.5 -- 1.6  \\
1.0  & ---  & 1.0 -- 1.1  \\
\hline
\end{tabular}
\end{center}
\end{table}

Figures 9 and 10 present a detailed study of the Hill estimator for samples of size $N = 10^6$
and $10^8$. Since for $\alpha\le 1.5$ the results were satisfactory even for much smaller 
samples, in this study we restricted ourselves to $\alpha=1.95$ and $1.8$.
When looking at the left panels of Fig. 9, i.e. for $k<10\%$ of the sample size (as in Fig. 8),
the flat regions suggest similar values of $\alpha_{Hill}$ as did the Hill plots in Fig. 8.
However, when we enlarge the pictures and plot the Hill statistics only for $k<0.1\%$ of 
the sample size we can observe a much better fit. For the $1.8$-stable sample Hill estimates 
are very close to the true value of $\alpha$ for $k\in (150,400)$. But for the "hopeless" case 
of $\alpha=1.95$ again there is no single value of $k$ that gives the right value of the tail 
exponent. Yet this time the estimates are very close to the Levy-stable regime. Like for 
smaller samples, the Hill estimates agree quite well with log-log regression estimates
of the true tail, see Figs. 3 and 5. 
For the extreme case of $10^8$ observations the results are similar. The flat regions suggest 
almost the same values of $\alpha_{Hill}$ as did the Hill plots in Fig. 9.
In fact the Hill plots for $k<10\%$ of the sample size look so much alike, that in Fig. 10 
we plotted only the counterparts of the right plots of Fig. 9, i.e. the enlarged plots for 
$k<0.1\%$ of the sample size. We can see that for $\alpha=1.8$ a good estimate is obtained 
only for $k=25000,...,60000$ (i.e. for $k<0.06\%$ of sample size), whereas for $\alpha=1.95$ the 
estimate is in the Levy-stable regime only for $k<20000$ (i.e. for $k<0.02\%$ of sample size).

\section{Conclusions}

In the previous Section we showed that widely used tail index estimates (log-log linear 
regression and Hill) can give exponents well above the asymptotic limit for Levy-stable 
distributions with $\alpha$ close to 2. As a result, tail indices are significantly 
overestimated in samples of typical size. Only very large data sets ($10^6$ observations 
or more) exhibit the true tail behavior and decay as $x^{-\alpha}$. In practice, this means 
that in order to estimate $\alpha$ we must use high-frequency asset returns and analyze 
only the most outlying values. Otherwise, inference of the tail index may be strongly 
misleading and rejection of the Levy-stable regime unfounded. 

Recently extremely large data sets have become available to the researchers. The largest 
ones studied in the literature so far (in the context of tail behavior) are 
(i) the 40 million data points record of 5 minute increments for 1000 U.S. companies during 
the two year period 1994-95 \cite{gmas98,pgams99} and 
(ii) the 1 million data points record of 1 minute increments for the Standard{\&}Poor's 500 
index during the 13-year period 1984-96 \cite{gmas98,gpams99}. 
However, the former data set is not homogeneous. It is formed out of 1000 data sets of 
40000 records for individual companies. Therefore strong correlations, which can conceal 
the true nature of asset returns, are present in the data. The estimated tail indices for 
individual companies (see Fig. 1(b) in Ref. \cite{pgams99}) were found to range from 
$\alpha=1.5$ to 5.5. As we have shown in the previous Section these values can be easily 
obtained for samples of Levy-stable distributed variables.

On the other hand, the second (ii) data set may be regarded as homogeneous \cite{SP500}. 
The tails of the distribution of 1 minute returns were reported to decay in a power-law
fashion with an exponent $\alpha=2.95$ for positive and $\alpha=2.75$ for negative 
observations \cite{gpams99}. In an earlier paper by the same authors \cite{gmas98} 
the tail exponents were reported to be 2.93 and 3.02, respectively, which shows that  
the log-log linear regression method is very sensitive to the choice of observations used.
Moreover, in both papers the tails of the empirical cumulative distribution function curl 
upward for extreme returns, see Fig. 1(c) in \cite{gmas98} and Fig. 4(a) in \cite{gpams99}. 
This suggests that for very large and very small observations the distribution could be 
fitted by a power-law with a much smaller exponent. For example, if in Fig. 4(a) of 
\cite{gpams99} we plot an approximate regression line for negative returns in the region 
$20\le g \le 100$ we find that the power-law exponent is less then 2. This may indicate
that the tail exponents reported in both papers did not refer to the true tail behavior, 
but to the initial power-like decay (see Figs. 3 and 5) and that the rejection of the 
Levy-stable regime was not fully justified.

In this paper we have shown that the reported estimated tail exponent around 3 
may very well indicate a Levy-stable distribution with $\alpha\approx 1.8$. 
This is consistent with earlier findings (for a review see \cite{rm00}) 
where the returns of numerous financial assets (individual stocks, indices, 
FX rates, etc.) were reported to be Levy-stable distributed with $\alpha\in (1.65,2]$. 
However, nothing we have said demonstrates that asset returns are 
indeed Levy-stably distributed. Although the analyzed tail index estimates are not 
sufficient to reject Levy-stability, by no means can we rule out a leptokurtic 
non-Levy-stable distribution that has power-law tails with $\alpha>2$.

\section{Acknowledgments}

The paper greatly benefited from Dietrich Stauffer's critical reading and helpful comments.
The research of the author was partially supported by KBN Grant no. PBZ 16/P03/99.


\begin{thebibliography}{99}
\vspace{-.3cm} \bibitem{STABLE} In the mathematical literature Levy-stable laws are called 
$\alpha$-stable or just stable. Such a name has been assigned to these distributions because 
a sum of two independent random variables having a Levy-stable distribution with index $\alpha$ 
is again Levy-stable with the same index $\alpha$. However, this invariance property does not
hold for different $\alpha$'s, i.e. a sum of two independent Levy-stable random variables with 
different tail exponents is not Levy-stable. 
\vspace{-.3cm} \bibitem{feller71} W. Feller, An Introduction to Probability Theory and its Applications, 2nd ed., Wiley, New York, 1971.
\vspace{-.3cm} \bibitem{zolotariev86} A. Zolotariev, One--Dimensional Stable Distributions, American Mathematical Society, Providence, 1986.
\vspace{-.3cm} \bibitem{jw94} A. Janicki, A. Weron, Simulation and Chaotic Behavior of $\alpha$-Stable Stochastic Processes, Marcel Dekker, New York, 1994.
\vspace{-.3cm} \bibitem{CLT} Recall that the classical Central Limit Theorem states that the limit 
of normalized sums of independent identically distributed terms with finite variance is Gaussian 
(Levy-stable with $\alpha=2$). 
\vspace{-.3cm} \bibitem{ekm97} P. Embrechts, C. Kluppelberg, T. Mikosch, Modelling Extremal Events for Insurance and Finance, Springer, 1997.
\vspace{-.3cm} \bibitem{rm00} S. Rachev, S. Mittnik, Stable Paretian Models in Finance, Wiley, 2000.
\vspace{-.3cm} \bibitem{dumouchel83} W.H. DuMouchel, Ann. Statist. 11 (1983) 1019.
\vspace{-.3cm} \bibitem{hv91} M.C.A.B. Hols, C.G. de Vries, J. App. Econometrics 6 (1991) 287.
\vspace{-.3cm} \bibitem{lp94} M. Loretan, P.C.B. Phillips, J. Empirical Finance 1 (1994) 221.
\vspace{-.3cm} \bibitem{longin96} F.M. Longin, J. Business 69 (1996) 383.
\vspace{-.3cm} \bibitem{dv97} J. Danielsson, C.G. de Vries, J. Empirical Finance 4 (1997) 241. 
\vspace{-.3cm} \bibitem{gmas98} P. Gopikrishnan, M. Meyer, L.A.N. Amaral, H.E. Stanley, Euro. Phys. J. B 3 (1998) 139. 
\vspace{-.3cm} \bibitem{gpams99} P. Gopikrishnan, V. Plerou, L.A.N. Amaral, M. Meyer, H.E. Stanley, Phys. Rev. E 60 (1999) 5305. 
\vspace{-.3cm} \bibitem{pgams99} V. Plerou, P. Gopikrishnan, L.A.N. Amaral, M. Meyer, H.E. Stanley, Phys. Rev. E 60 (1999) 6519. 
\vspace{-.3cm} \bibitem{pgras00} V. Plerou, P. Gopikrishnan, B. Rosenow, L.A.N. Amaral, H.E. Stanley, Physica A 279 (2000) 443. 
\vspace{-.3cm} \bibitem{la01} T. Lux, M. Ausloos, Market Fluctuations I: Multi-Scaling and Their Possible Origins, in A. Bunde, H.-J. Schellnhuber eds., Theories of Disasters, Springer, 2001. 
\vspace{-.3cm} \bibitem{levy25} P. Levy, Calcul des Probabilites, Gauthier Villars, Paris, 1925.
\vspace{-.3cm} \bibitem{nolan97} J.P. Nolan, Comm. in Statist. -- Stochast. Models 13 (1997) 759. The computer program STABLE can be downloaded from http://academic2.american.edu/$\tilde{~}$jpnolan/stable/stable.html.
\vspace{-.3cm} \bibitem{hall81} P. Hall, Bull. London Math. Soc. 13 (1981) 23.
\vspace{-.3cm} \bibitem{st94} G. Samorodnitsky, M.S. Taqqu, Stable Non--Gaussian Random Processes, Chapman \& Hall, New York, 1994. 
\vspace{-.3cm} \bibitem{weron96} R. Weron, Statist. Probab. Lett. 28 (1996) 165. See also: R. Weron, "Correction to: On the Chambers-Mallows-Stuck method for simulating skewed stable random variables", Research Report, Wroc{\l}aw University of Technology, 1996, http://www.im.pwr.wroc.pl/$\tilde{~}$hugo/Publications.html.
\vspace{-.3cm} \bibitem{fn99} H. Fofack, J.P. Nolan, Extremes 2 (1999) 39.
\vspace{-.3cm} \bibitem{kanter75} M. Kanter, Ann. Probab. 3 (1975) 697. 
\vspace{-.3cm} \bibitem{cms76} J.M. Chambers, C.L. Mallows, B.W. Stuck, J. Amer. Statist. Assoc. 71 (1976) 340.
\vspace{-.3cm} \bibitem{dumouchel73} W.H. DuMouchel, Ann. Statist. 1 (1973) 948.
\vspace{-.3cm} \bibitem{nolan99} J.P. Nolan, "Maximum likelihood estimation of stable parameters", Preprint, American University. 
\vspace{-.3cm} \bibitem{koutrouvelis80} I.A. Koutrouvelis, J. Amer. Statist. Assoc. 75 (1980) 918.
\vspace{-.3cm} \bibitem{koutrouvelis81} I.A. Koutrouvelis, Commun. Statist.--Simula. 10 (1981) 17.
\vspace{-.3cm} \bibitem{kw98} S.M. Kogon, D.B. Williams, Characteristic function based estimation of stable parameters, in: R. Adler, R. Feldman, M. Taqqu, eds., A Practical Guide to Heavy Tails, Birkhauser, Boston, 1998.  
\vspace{-.3cm} \bibitem{mcculloch86} J.H. McCulloch, Commun. Statist.--Simula. 15 (1986) 1109. 
\vspace{-.3cm} \bibitem{hill75} B.M. Hill, Ann. Statist. 3 (1975) 1163.
\vspace{-.3cm} \bibitem{HILL} The method can be applied to the lower tail of a distribution as 
well. Usually it is enough to multiply the sample by $-1$ and then proceed as for the upper tail.
\vspace{-.3cm} \bibitem{resnick97} S. Resnick, Ann. Statist. 25 (1997) 1805. 
\vspace{-.3cm} \bibitem{bvt96} J. Beirlant, P Vynckier, J.L. Teugels, J. Amer. Statist. Assoc. 91 (1996) 1659.
\vspace{-.3cm} \bibitem{dk98} H. Drees, E. Kaufman, Stochast. Process. Appl. 75 (1998) 149.
\vspace{-.3cm} \bibitem{mcculloch97} J.H. McCulloch, J. Business Econ. Statist. 15 (1997) 74. 
\vspace{-.3cm} \bibitem{rt97} R.-D. Reiss, M. Thomas, Statistical Analysis of Extreme Values, Birkhauser, Boston, 1997. 
\vspace{-.3cm} \bibitem{SP500} However, during the 13-year period many things have changed in 
the financial markets. For instance, the trading volume and the speed of information arrival 
have increased enormously. 

\end{thebibliography}
\end{document}